# Imaging artefacts in scattering-type scanning near field optical microscopy arising from optical diffraction effects and contrast-active sub-surface features

Denis E. Tranca[1], Stefan G. Stanciu[1], Radu Hristu[1], Yotam Schatzberg[2], Zeev Zalevsky[2*], Binyamin Kusnetz[3], Avi Karsenty[3], Cosmin K. Banica[4], George A. Stanciu[1*]

[1] Center for Microscopy-Microanalysis and Information Processing, National University of Science and Technology Politehnica Bucharest
[2] Faculty of Engineering, Bar-Ilan University (BIU), Ramat Gan 5290002, Israel
[3] Faculty of Engineering, Jerusalem College of Technology, Jerusalem 9116001, Israel
[4] SC Wing Computer Group SRL, Bucharest, Romania
* stanciu@physics.pub.ro; zeev.zalevsky@biu.ac.il

**Abstract**

The scattering-type Scanning Near-Field Optical Microscope (s-SNOM) is acknowledged as an excellent tool to investigate the optical properties of different materials and biological samples at the nanoscale. In this study we show that s-SNOM data are susceptible to being affected by specific artefacts related to the light diffraction phenomena and to stray contributions from shallow buried, contrast-active, structures. We focus on discussing the diffraction contributions from sample edges, next to those corresponding to one- or two-dimensional periodic structures, and undesired contributions from shallow buried periodic features. Each scenario was examined individually through both experimental methods and simulations. Our experimental findings reveal that such artefacts affect not only s-SNOM images demodulated at the direct-current (DC) component and the fundamental frequency, but also images demodulated at higher harmonic frequencies. We show that image artefacts caused by diffraction resemble the undesirable effects caused by illumination with a laser beam of unstable intensity, and that buried features can yield s-SNOM signals that cannot be distinguished from those originating from the sample surface, in absence of prior knowledge of the sample structure. Performed simulations confirm these experimental findings. This work enhances the understanding of s-SNOM data and paves the way for new data acquisition and postprocessing methods that can enable next-generation s-SNOM imaging and spectroscopy with significantly enhanced signal-to-noise ratio and resolution.

## 1. Introduction

Among super-resolved microscopy techniques providing optical resolutions beyond the diffraction limit, scattering-type scanning near-field optical microscopy (s-SNOM) stands out as one of the most effective techniques for nanoscale surface imaging, capable of producing images with lateral optical resolutions of approximately 10 nm, and below in special settings [1] while providing chemical contrast[2,3]. These capabilities make s-SNOM an excellent, and in some cases irreplaceable tool for a wide range of applications in nanotechnology, chemistry, materials science, life sciences and other fields[4–8]. Traditionally, an s-SNOM system integrates an atomic force microscope (AFM) with the key element enabling nanoscale optical imaging being a laser beam focused on the tip of the probe, usually from the lateral direction. The light scattered by the tip during surface scanning is modulated by the sample's optical parameters, resulting in a pair of amplitude and phase images [9,10], with amplitude signals being attributed to the reflectivity of the sample, and the phase signals being correlated with its absorption [11]. The reconstruction of the amplitude and phase signals is not straightforward, given the tiny volume around the tip apex contributing to the signals of interest, compared with the intense background light originating from neighboring areas. Usually, s-SNOM signal reconstruction requires complex detection schemes, with interferometric detection and higher harmonic demodulation being the most common [9]. Such detection strategies have the main purpose to increase the signal-to-noise ratio (SNR), and thus enable the visualization of faint near-field signals originating from nanoscale features on the sample surface, or from shallow buried features, as reported in a few studies [12,13]. The primary sources of noise are topography crosstalk [14] and background radiation [9,15], which typically originates from multiple reflections of the incident light between the probe and the sample surface [16], finally reaching the detector and producing image artefacts. Fortunately, demodulation of the detected signal at higher harmonics, in conjunction with interferometric detection schemes, such as pseudo-heterodyne detection [9], performs well in consistently reducing, and in some cases, completely eliminating the unwanted contribution of the background radiation.

However, there are still additional image artefacts caused by other effects occurring due to sample's physical and optical characteristics, which have been sparsely documented to date, although some of them are well known to experienced s-SNOM practitioners. In this article we focus on this issue, discussing for the first time to our knowledge, artefacts arising from diffraction effects occurring at the edges/boundaries of imaged elements on the investigated sample, as well as from standard diffraction effects caused by one-dimensional (1D) and two-dimensional (2D) periodic elements acting as diffraction gratings. Additionally, we experimentally demonstrate, model and discuss artefacts generated by shallow buried, contrast active features.

The occurrence of interference and diffraction patterns at the sample surface is not unusual in optical microscopy [17–19], but to the best of our knowledge has not been discussed to date in the context of s-SNOM imaging. However, we argue that in particular cases when the investigated sample contains boundaries or periodic elements of materials with high differences in optical reflectivity and transmission, the effects of diffraction are pronounced, and misleading, being an important source of potential data interpretation flaws. In this article, besides highlighting the existence of such artefacts and their impact on s-SNOM imaging, we advance the hypothesis that when a diffraction pattern is formed at the surface of the sample, the scattered near-field light is also modulated by the light intensity of the optical diffraction pattern. In typical s-SNOM models, the scattered near-field light is proportional to the local effective polarizability (dependent on the local properties of the sample) and to the incident light intensity which is maintained constant. When the incident light intensity fluctuates, the image is highly affected. We hypothesize that a diffraction pattern localized at the surface of the sample behaves like an incident light with fluctuant intensity, influencing the amplitude and phase of the scattered light and producing image artefacts. Part of the work presented is focused on demonstrating this, by experimental means and by simulations.

Another aspect tackled by our study refers to in-depth s-SNOM detection, which although an important capability of this imaging technique, has been reported only in a limited number of studies [13,20–25]. As could be expected, due to the required tip-sample interactions, the axial resolution of s-SNOM is limited, and compatible with the detection of shallow buried features, up to a depth of around 100 nm [13]. Irrespective of this limitation, s-SNOM's capabilities for sub-surface imaging are regarded as highly useful given the importance of resolving nanosized sub-surface features and structures, especially in the case of emerging nanomaterials and nanodevices [12,13,20]. However, a side-effect of this capability is that sub-surface features with different thicknesses and different distances to the top surface will influence the image contrast differently. When positioned on a top of a substrate, the thickness of a thin sample of interest, e.g. a flake, can be measured by AFM, and this information can be correlated to the s-SNOM analysis to aid data interpretation. However, neither AFM nor s-SNOM have the possibility to detect the thickness or the depth of buried features without prior knowledge about the sample under investigation. Therefore, AFM and in-depth s-SNOM detection are unable to resolve the exact depth or thickness of buried features. One of the main issues caused by buried features is the parasitic contrast arising from such sub-surface, contrast-active, structures in the sample, of which the s-SNOM practitioner performing the imaging session might not be aware. To this end, we refer to the case of a thin film (e.g. <50 nm) of homogenous composition and contrast, which should yield an s-SNOM image of homogenous contrast. Supposing that the thin film is positioned on top of a substrate contaminated with a contrast-active material, the recorded s-SNOM images will add on top of the s-SNOM signals corresponding to the thin film, parasitic s-SNOM signals from the buried contrast-active materials, resulting in inhomogeneous s-SNOM images that raise significant interpretation problems. Furthermore, when the buried features form a diffraction grating, then the effects may be even worse as diffraction artefacts will add up on top of buried-features-related artefacts. We experimentally demonstrate the existence of such artefacts, which we confirm by simulations. Although not addressed in our work, such situations related to parasitic s-SNOM contrast of buried features can be extrapolated also to the case of nanoparticle imaging [26], where we can consider for instance the case of nanoparticles possessing contrast-active contaminants at the core, which can bias the s-SNOM measurements on the considered nanoparticle.

After presenting the investigated samples, imaging tools and simulation models, we discuss the following cases: (i) artefacts occurring due to edge diffraction; (ii) artefacts yielded by 1D and 2D periodic structures; (iii) artefacts occurring due to parasitic contrast arising from buried features. All cases are demonstrated and documented by experimental measurements and simulation methods. Overall, this work contributes to the better understanding of s-SNOM data and paves the way for new data acquisition and postprocessing methods that can enable next-generation s-SNOM imaging and spectroscopy with significantly enhanced SNR.

## 2. Materials and methods

### 2.1. Samples

**Edge diffraction:** the sample used for studying the edge diffraction effects consists of a thin Ag layer (~100 nm thickness) deposited on a BK7 glass substrate by thermal evaporation in a VU-2M vacuum equipment at a deposition rate of 5 Å/sec. After deposition, part of the Ag layer was mechanically removed using a sharp needle to obtain an axial edge between the Ag film and glass substrate. Top-view and cross-section schematics of this sample are represented in Fig. 1.a).

**Diffraction from 1D periodic structures:** the samples used for studying the effects arising from optical diffraction on 1D periodic structure was a standard calibration sample, typically commercialized by Park Systems (Korea) for electric force microscopy (EFM) calibration. It consists of two interdigitated microcomb-shaped Au electrodes structures deposited on a Si substrate. Because of its periodic micro-structure, this sample also acts as a diffractive grating in the reflective regime and this property is used in our study to investigate s-SNOM artefacts arising from diffraction on 1D periodic structures. The top-view and cross-section schematics of this sample are represented in Fig. 1.b).

**Diffraction from 2D periodic structures and buried features:** For the studies performed on s-SNOM artefacts arising from optical diffraction on 2D periodic structures and contrast-active buried features we used three types of samples extensively described and studied in our previous work [27]. In brief, these consist of three configurations, Fig. 1.c)-e):

- Au periodic square flakes (1x1 $\mu m^2$ area with a thickness of 10 nm) deposited on the surface of a thin $SiO_2$ thin film (40 nm thickness), which itself is deposited on a Si substrate (configuration A) – Fig. 1.c);

- Au circular periodic flakes (1 μm in diameter and 10 nm thick) deposited directly on top of a Si substrate and covered by a 40 nm $SiO_2$ thin film (configuration B) – Fig. 1.d);

- Au circular periodic flakes (1 μm in diameter and 10 nm thick) deposited on the Si substrate, covered by a 40 nm $SiO_2$ thin film, on top of which Au square periodic flakes (1x1 $\mu m^2$ area with a thickness of 10 nm) are deposited (configuration C). This last configuration represents a combination of the first two, A and B – Fig. 1.e).

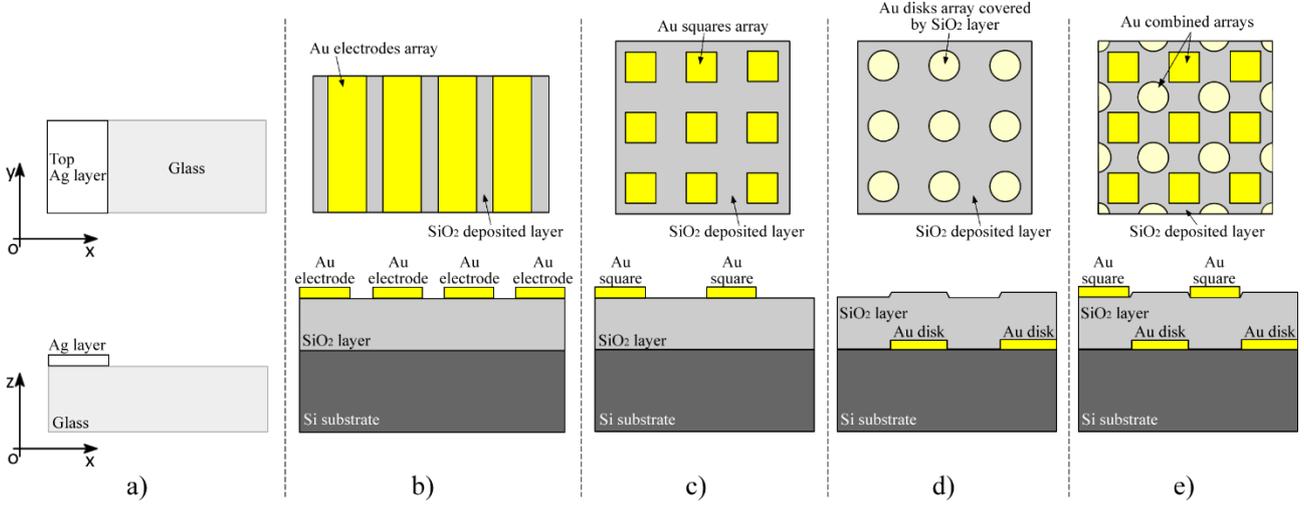

*Figure 1*. Schematic top-view (xOy plane) and cross-section (xOz plane) details of the samples used for studying: a) edge diffraction artefacts in s-SNOM; b) diffraction artefacts in s-SNOM generated by 1D periodic structures; c) 2D periodic structure on top of the surface, configuration A; d) buried 2D periodic structures, configuration B; e) top and buried 2D periodic structures, configuration C.

## 2.2. Imaging tools

**Edge diffraction:** For studying edge diffraction artefacts we used a previously reported home-built multimodal prototype system for far-field and near-field microscopy [28,29] which incorporates an s-SNOM system, mounted on top of an inverted optical microscope, providing possibilities for illuminating the s-SNOM tip from the lateral side, as in conventional s-SNOM systems, but also from beneath the sample, via the optical path enabled by the inverted optical microscope. This s-SNOM configuration was preferred here given its similarities with the typical optical setups used for studying optical diffraction near sharp edges [30]. The probe tip was illuminated by means of a supercontinuum kit (Thorlabs, SCKB2) pumped by a Ti:Sapphire fs pulsed laser (Chameleon Vision II, Coherent), characterized by 140 fs pulse width, 80 MHz repetition rate and wavelength set to 800 nm. The wavelength of the beam from the supercontinuum source was selected using the Super K Select Multiline Filter (NKT Photonics), equipped with an AOTF tunable in the range of 450-700 nm. For wavelengths above 700 nm, we used the laser beam directly from the Ti:Sapphire laser.

**Diffraction from periodic structures and buried features:** To investigate the other types of artefacts, we used a commercial s-SNOM (Nea-SNOM, Neaspec), working in a pseudo-heterodyne detection configuration with lateral tip illumination. The illumination beam used in this imaging configuration was a continuous-wave near-infrared laser working at 1550 nm (DFB Pro laser diode, Toptica).

For both s-SNOM systems we used an Au coated probe (Mikromasch, HQ:NSC18/Cr-Au), with a nominal oscillation frequency of 75 kHz and a nominal force constant of 2.8 N/m. Since s-SNOM systems inherently include AFM systems operating in tapping mode, topography images were automatically acquired along with the s-SNOM images.

## 3. Simulation methodology and theoretical framework

The AFM and s-SNOM images were post-processed using Gwyddion software[31], which was employed for data leveling and background subtraction.

The MATLAB® software was used for both s-SNOM data analysis and simulation. The goal of the MATLAB® simulations was to verify the hypothesis that optical intensity of the diffraction pattern affects the s-SNOM images even for the case of higher harmonic demodulation. The simulated s-SNOM images were obtained according to a previous methodology introduced in[32–34] which makes use of the oscillating point-dipole (OPD) model[35]. We opted to use the OPD over the finite-dipole (FD) [36] model because it is mathematically less complex and computationally faster. For each discussed case, the topography image and sample design characteristics were used to create an image mask, which was then employed to segment the sample areas corresponding to different materials. This mask and segmentation were subsequently utilized to generate a virtual sample, represented as a multidimensional array containing physical characteristics like sizes and positions, reflecting the properties of the real sample. Each element of the array also includes information on the sample's optical characteristics, specifically its complex electric permittivity. By applying an adapted version of the simulations from [32,34] we modeled the s-SNOM amplitude and phase signals for each array element, generating simulated s-SNOM images for several harmonics.

The MATLAB® simulations were adapted to consider the tip illuminated not only by a light source of constant intensity, but also by an additional light intensity resulting from diffraction, which varies spatially from one point to another. This intensity pattern caused by diffraction is modeled separately by different means depending on the specific case, as will be discussed in the following subsections.

### 3.1. Intensity pattern caused by diffraction for the case of edge diffraction

The theory of light diffraction at sharp edges was established in the past [37–39], and recent studies have introduced alternative models and applications of this effect [30]. For the sake of conciseness, we will not reproduce in this work the theoretical and mathematical details of the edge diffraction phenomena, but we find it important to recall that the basic parameters involved in this effect are the wavelength of the light, the average distance from the sample surface to the probe tip, and the angle of incidence, which varies in the limit of the numerical aperture of the microscope objective (NA = 0.3, in the case of our experiment). We adapted in MATLAB® the model presented in [40] to obtain the diffraction intensity for this case. While incidence at a certain angle produces an intensity variation with unevenly spaced fringes, it was observed that the superposition of multiple waves at different angles (within the numerical aperture limit) produces an intensity variation that more closely resemble equally spaced fringes, which is in accordance with the experimental results.

After simulating the diffraction intensity variation due to edge diffraction, the next step is to build the intensity pattern specific to this case. An AFM topography image was used to create a mask for segmenting the sample areas corresponding to different materials. Because the illumination and the detection are done on opposite sides of the sample (due to transmission configuration), the diffraction pattern occurs over the transparent part of the sample (on the top plane of the BK7 glass substrate). The resulting diffraction pattern was further used in the MATLAB® s-SNOM model to get the simulated amplitude s-SNOM image demodulated on 3$^{rd}$ harmonic.

Simulation results and comparison with the experimental case are presented in the Results section, for different wavelengths in the visible spectrum.

### 3.2. Intensity pattern caused by diffraction for the case of 1D grating

For the case of the 1D grating the diffraction pattern was obtained from the direct current (DC) component of the detected s-SNOM signal. It is well-known that in the DC component, the near-field, high-resolution details specific to s-SNOM are heavily affected by the presence of the optical background noise [9]. Among other types of noise (like topography-related noise and far-field reflections), the background noise contains far-field diffraction-related noise, which in this case is basically the diffraction pattern from the 1D grating. To obtain the diffraction pattern, the key is to remove the high-resolution details from the DC image. This can be done by filtering out the high-frequency components using a Fast Fourier Transform (FFT) filtering, which can be achieved with the Gwyddion software. The resulting diffraction pattern together with topography, the experimental s-SNOM amplitude image and the simulated s-SNOM image are presented in the Results section.

### 3.3. Intensity pattern caused by diffraction for the case of 2D gratings

In tandem with the experimental approach, a numerical analysis of the fabricated samples containing 2D gratings was applied. Ansys High-Frequency Structure Simulator (HFSS) was used to simulate the full 3D electromagnetic (EM) field of samples with 2D gratings. HFSS is a popular simulation tool used in the field of electromagnetic design and analysis [41]. It employs the finite element method (FEM) to solve Maxwell's equations, allowing us to design and optimize complex electromagnetic and high-speed electronic devices.

For these simulations, an HFSS model was constructed for each of the three samples containing 2D gratings (configurations A, B and C already presented in the Samples section).

Each simulated sample model is comprised of a 3×3 unit cell. This is a compromise between our aim to achieve high resolution simulation results, minimize edge field effects and the computational power currently available to us.

Next, a radiation source was defined. In an attempt to simulate the s-SNOM (Nea-SNOM, Neaspec) device, a lateral illuminating plane wave with a wavelength of 1550 nm was selected (beam diameter = 6 μm, lateral beam at 45 deg relative to the sample XY plane). The simulation boundary conditions were set to infinitely radiating volume, to prevent boundary reflections of radiation in the simulation calculation.

The model mesh element size was set to be less than one-sixth of the wavelength. The resulting images represent the magnitude of the electric field on the surface of the model. The intensity pattern caused by diffraction was obtained by calculating the square of the electric field's magnitude and normalizing the result to unit.

### *3.4. Simulation of s-SNOM artefacts arising from parasitic detection of sub-surface features*

s-SNOM's capabilities to detect signals arising from sub-surface structures were presented in several studies performed to date[12,13,20]. While the theoretical explanation of in-depth detection is still under development, advancements in the field concluded so far that s-SNOM's ability to detect signals arising from buried elements is linked to the near-field interaction region which includes contributions from sub-surface features [42,43]. Here we propose a theoretical model for in-depth s-SNOM detection, based on which we obtain simulated results that align with the experimental s-SNOM data. For this purpose, we designed 3D arrays emulating the real samples largely described in[27] and illustrated in Fig. 1.d),e) - configurations B and C.

The existing theoretical models (like the OPD and others) do not consider the sample thickness but focus solely on the sample material (via its dielectric function, $\varepsilon$), and the distance $d$ from the probe tip to the sample surface. In reality, however, a sample may consist of multiple parts (e.g. stacked layers) across the Z axis (height), each with different thicknesses and positioned at a different depth.

In the MATLAB® simulations the samples were considered as consisting of a multitude of planar thin layers, each layer having a certain spatial distribution of the two materials (Au and $SiO_2$), depending on the sample type, material depth and the shape of the features. For each planar layer a thickness of 1 nm was considered in the simulations.

For each such layer we calculated the s-SNOM amplitude and phase based on the OPD theoretical model, adjusting the tip-sample distance $d$ according to the depth of the layer (measured from the tip position).

For every $(x, y)$ coordinate on the sample surface plane, the total effective polarizability ($\alpha_{eff}$) was calculated as the sum of the effective polarizabilities obtained for each depth (on $z$ axis):

$$\alpha_{eff}(x,y) = \sum_{z=1}^{N} \alpha_{eff}(x,y,h_z) \qquad (1)$$

where $N$ is the number of total layers (100, in our case), and $h_z$ is the distance between the tip and the $z^{th}$ layer of the sample (counting from the top surface).

Using an adapted version of the simulations described in [32,34] in which we integrated the above formula for calculating the effective polarizability, we were able to simulate s-SNOM amplitude images for the first 5 harmonics of the probe oscillation frequency.

## 4. Results and discussions

### 4.1. Image artefacts arising from edge diffraction

Results on edge diffraction artefacts in s-SNOM imaging are displayed in Fig. 2. Fig. 2.a) shows the topography of the sample and Fig. 2.b) shows the binary mask used for creating simulation conditions similar to the experiment. In Fig. 2.c) we represent the experimental s-SNOM 3rd harmonic amplitude images for three different laser beam wavelengths: 550 nm, 635 nm and 800 nm, respectively. Fig. 2.d) displays the corresponding simulated images for the same wavelengths. Comparing the experimental results to the simulations, we observe two important similarities. First, the average fringe distance increases with wavelength in both experiment and simulation. Second, the light intensity decreases with distance from the edge. Fig. 2.e) highlights these similarities by representing the Fast Fourier-Transform (FFT) of profiles over the dotted lines in Fig. 2.c) and 2.d) orthogonal on the edge. FFT evaluation reveals a clear consistency between experiments and simulations in respect to peak position and peak displacement as a function of wavelength.

The differences between experimental and simulated results can be attributed to the approximations used in the simulation. For instance, in close proximity to the edge the phase difference between adjacent points on the wavefront becomes sensitive to the exact curvature of the wavefront, which is not fully accounted for in the simulation model.

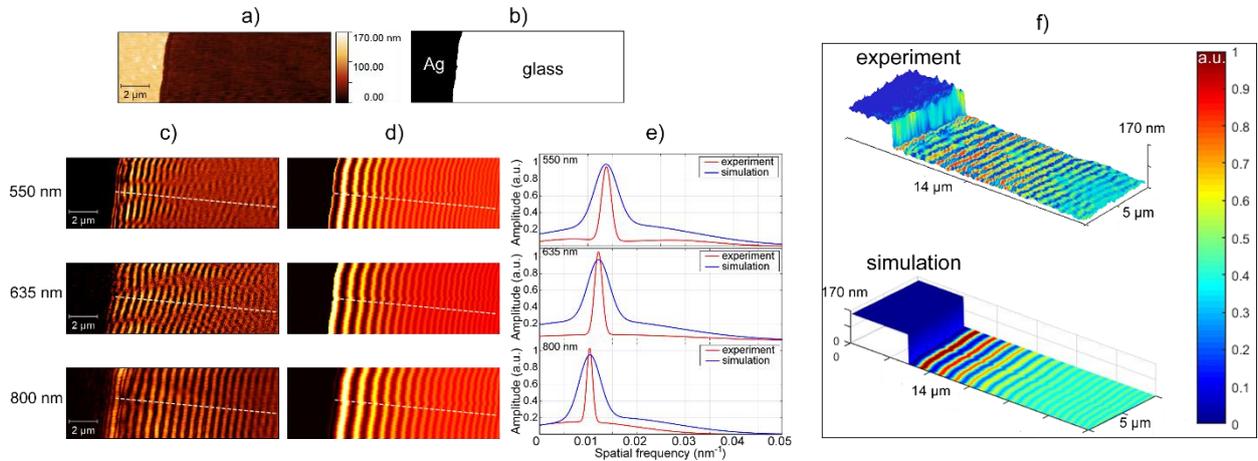

*Figure 2.* Edge diffraction study; a) AFM topography of the sample; b) Binary mask obtained based on the topography; c) s-SNOM 3rd harmonic amplitude images experimentally acquired for three different wavelengths (550 nm, 635 nm and 800 nm); d) Simulated s-SNOM amplitude images obtained for the same wavelengths; e) Fast Fourier-transform of profile lines in figures 1.c) and 1.d) showing the dependency of peak spatial period on the wavelength, for both experimental and simulated images; f) overlay image between 3D topography and color-coded s-SNOM amplitude in experiment and simulation.

### 4.2. Image artefacts arising from diffraction effects associated with 1D gratings

The results on artefacts arising from diffraction effects from 1D diffraction grating are presented in Fig 3.

In the topography image (Fig. 3.a)) the periodicity of the Au micro-electrodes forming the diffractive structure can be observed. Fig. 3.b) represents the 3rd harmonic s-SNOM amplitude image, where it can be seen that the amplitude value over the Au electrodes is not uniform but in fact shaped by the diffraction pattern.

Fig. 3.c) represents the DC component of the s-SNOM signal, which contains background radiation, other noises, and the periodical pattern characteristic to optical diffraction. After removing the high-frequency noise we get the image presented in Fig. 3.d), which contains mainly the optical diffraction pattern. It can be observed that the fringes are parallel to the electrodes as expected, but not completely overlapped to the these because the sample is illuminated laterally from a certain angle, as usual in the case of conventional s-SNOM imaging.

The simulated 3$^{rd}$ harmonic s-SNOM amplitude image is represented in Fig. 3.e). For this simulation we considered the sample illuminated by a constant intensity incident laser and by the diffraction pattern from Fig. 3.d). The similarity between experiment (Fig. 3.b)) and simulation (Fig. 3.e)) is clear, as both show higher intensity over certain areas of the Au electrodes and lower intensity over others. The images are rotated at a small angle to get the direction of the electrodes vertical for easier MATLAB® software processing.

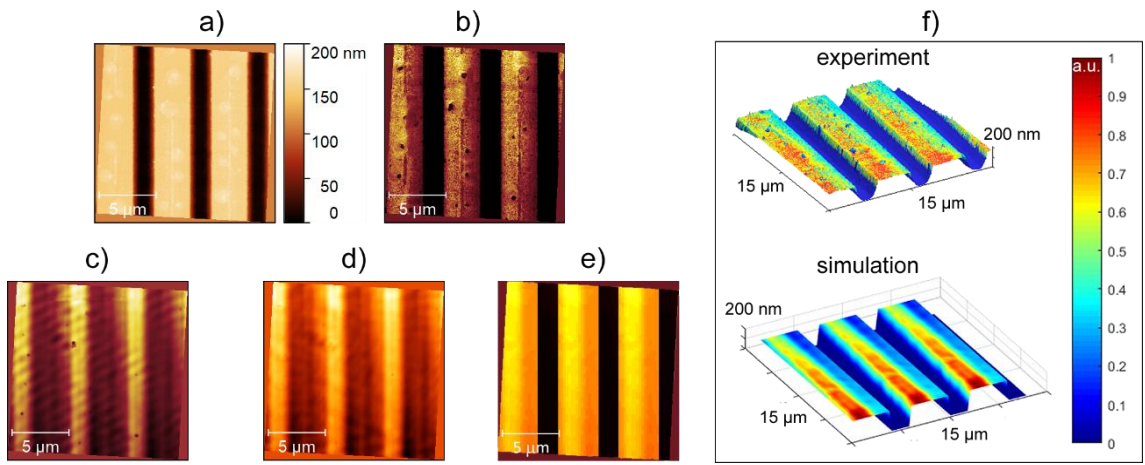

*Figure 3.* Experimental investigations and simulation on sample harboring 1D diffractive grating; a) AFM topography of the sample; b) 3$^{rd}$ harmonic s-SNOM amplitude image; c) DC component image of the s-SNOM signal; d) DC component image after removing high-frequency noise with FFT filtering; e) Simulated s-SNOM amplitude image; f) overlay image between 3D topography and color-coded s-SNOM amplitude in experiment and simulation.

### 4.3. Image artefacts arising from diffraction effects associated with 2D gratings

The results on artefacts arising from diffraction effects from a 2D diffraction grating consisting of square elements are presented in Fig 4. Fig. 4.a) presents the topography of the sample, highlighting the periodic elements of the grating. Fig. 4.b) presents the 3$^{rd}$ harmonic s-SNOM amplitude image. The magnitude of electric field on the surface of the sample simulated with HFSS is represented in Fig. 4.c). This image is used to get the diffraction-induced intensity pattern necessary for simulating the s-SNOM amplitude image. The simulated s-SNOM 3$^{rd}$ harmonic amplitude image is represented in Fig. 4.d). This image was cropped to illustrate a similar area as in the case of the experiment.

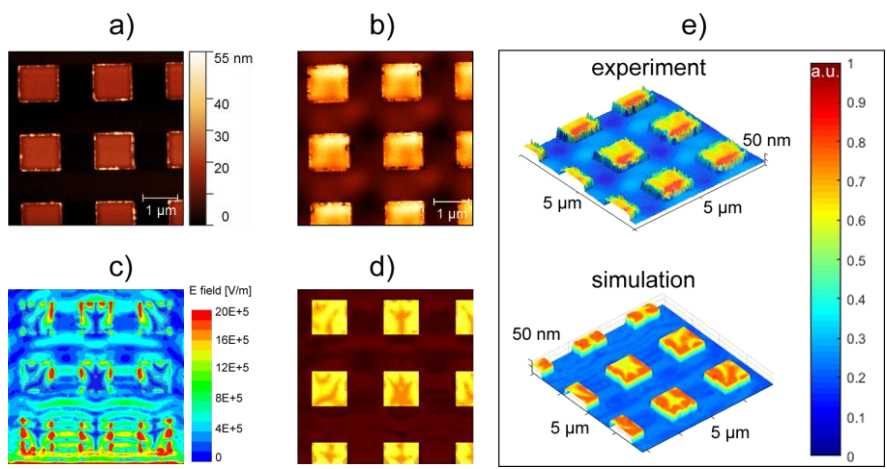

**Figure 4.** *Experimental investigations and simulation on a custom-made Si-SiO$_2$-Au sample (configuration A) harboring surface 2D diffractive gratings; a) AFM topography of the sample; b) 3$^{rd}$ harmonic s-SNOM amplitude image; c) Magnitude of electric field on the surface of the sample, simulated with HFSS; d) Simulated s-SNOM amplitude image; e) overlay image between 3D topography and color-coded s-SNOM amplitude in experiment and simulation.*

Here we can observe that the intensity distributed over the squares is nonuniform in both experimental and simulated s-SNOM images. Moreover, the intensity over the SiO$_2$ substrate fluctuates as well in both images.

### *4.4. Image artefacts arising from diffraction effects associated with sub-surface 2D gratings*

While the previous discussion was focused on surface positioned 2D gratings, we will now turn our attention to the case where the diffraction grating (made from disk-like flakes of Au) is positioned beneath the surface, buried under a thin SiO$_2$ layer. Studies performed on such samples having sub-surface contrast-active features are significant not only in terms of diffraction-related artefacts but also in the context of simulating in-depth s-SNOM detection, which to the best of our knowledge has not been demonstrated before. The topography image (Fig. 5.a) clearly reveals the locations of the disks, as the SiO$_2$ layer is not top-flat, but it follows the topography underneath. The profile of the sample is depicted in Fig. 1.d). For the 1550 nm wavelength used in the experiment the deposited SiO$_2$ thin layer is practically transparent, therefore the buried Au features located between SiO$_2$ layer and Si substrate still perform as a diffraction grating. In Fig. 5.b) we depict the experimental 3$^{rd}$ harmonic s-SNOM amplitude image. Fig. 5.c) represents the electric field's magnitude on the surface of the sample simulated with HFSS. Calculating the square of the values in this image we get the intensity pattern which influences the simulated 3$^{rd}$ harmonic s-SNOM amplitude image represented in Fig. 5d). Because the Au features are buried, the simulations were done by considering a typical detection depth of 100 nm.

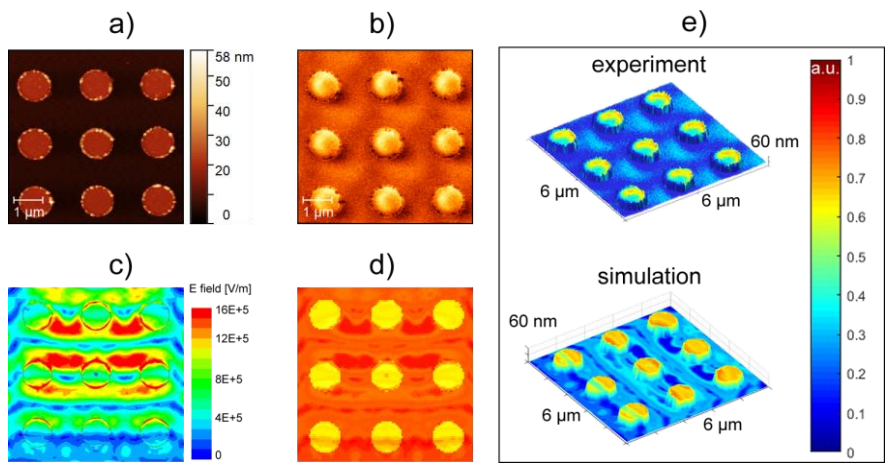

**Figure 5.** *Experimental investigations and simulation on a custom-made Si-Au-SiO$_2$ sample (configuration B) harboring sub-surface 2D diffractive gratings; a) AFM topography of the sample; b) 3$^{rd}$ harmonic s-SNOM amplitude image; c) Magnitude of electric field on the surface of the sample, simulated with HFSS; d) Simulated s-SNOM amplitude image; e) overlay image between 3D topography and color-coded s-SNOM amplitude in experiment and simulation.*

The experimental and simulated images displayed in Fig. 5.b) and 5.d) respectively, are highly similar. First, in both cases the s-SNOM amplitude over the location of the Au disks is non-uniform, and second, the s-SNOM amplitude over the rest of the surface (bare Si-SiO$_2$ sample regions) is non-uniform as well, showing periodic intensity maxima and minima.

These similarities consolidate the hypothesis that diffraction effects caused by the investigated sample can induce significant artefacts in s-SNOM images that can severely hinder data interpretation.

Besides the specific artefacts induced by the sub-surface diffraction gratings, it is important to note here that AFM and s-SNOM investigations cannot detect that the Au disks are in fact buried and not located on top of the sample. This information is a priori known to us, given the sample's geometry that we finely tuned upon custom, in-house, sample nanofabrication. However, in the potential case in which this information would have been unknown, the obtained AFM and s-SNOM images could have been easily misinterpreted to conclude that the Au disk structures are positioned on the surface on the sample, and not buried under the $SiO_2$ layer. An additional negative side-effect to such contrast arising from sub-surface elements, whose origin across the Z-axis cannot be distinguished, is represented by the fact that it can interfere with s-SNOM signals collected on the sample surface. Thus, an important conclusion is that s-SNOM images collected on structures positioned on the sample surface can be affected by parasitic signals corresponding to sub-surface, contrast-active, structures, which will overlap with the signals of surface features.

### *4.5. Image artefacts arising from diffraction effects associated with both top and sub-surface 2D gratings*

For this case study we used the sample described in the Methods section as configuration C, consisting of Au disk structures deposited on a Si substrate, which are covered by a $SiO_2$ layer, on top of which Au square structures are deposited. This sample is basically a combination of two 2D gratings, one located on top of the sample, and the other buried under the sample surface.

Here, we employed the same procedures as in the previous two cases. Accordingly, in Fig. 6.a) we present the AFM topography, in Fig. 6.b) the $3^{rd}$ harmonic s-SNOM amplitude image, in Fig 6.c) the electric field's magnitude on the surface of the model simulated with HFSS, and in Fig 6.d) the simulated s-SNOM image.

We observe that the experimental and simulated images for this sample configuration are quite similar and share many common features. The experimental s-SNOM amplitude over both squares and disks yields non-uniform profiles, as well as over the $SiO_2$ surface. A similar non-uniformity over the three regions is observed as well in the simulated s-SNOM image.

Furthermore, the overall intensity over the Au squares (which are located on top of the sample) is much higher compared to the intensity over the Au disks, which are located at about 40 nm depth under the surface. This means that the simulation succeeds at least qualitatively in modeling the sub-surface detection. In a simple quantitative analysis, the ratio between the average intensity corresponding to the surface-located Au squares and the average intensity corresponding to the buried Au disks in the experiment is comparable to the same ratio calculated for the simulation case.

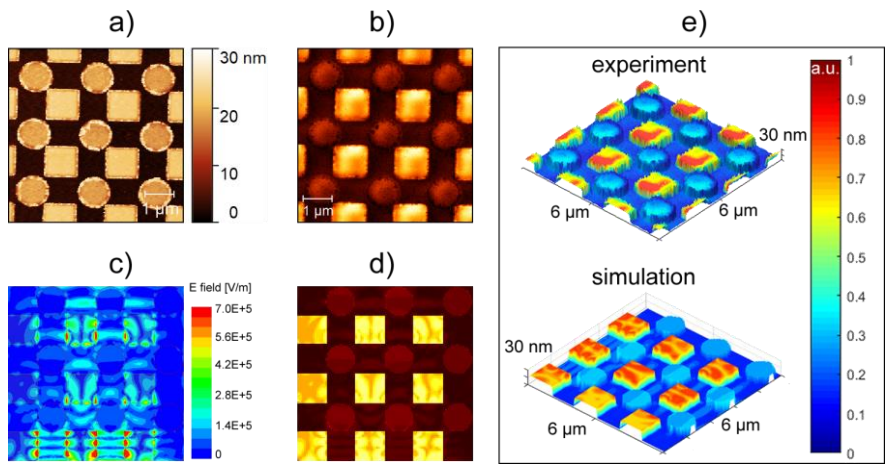

*Figure 6. Experimental investigations and simulation on a custom-made Si-Au-SiO$_2$-Au sample (configuration C) harboring surface and sub-surface 2-D diffractive gratings; a) AFM topography of the sample; b) $3^{rd}$ harmonic s-SNOM amplitude image; c) Magnitude of electric field on the surface of the*

*sample, simulated with HFSS; d) Simulated s-SNOM amplitude image; e) overlay image between 3D topography and color-coded s-SNOM amplitude in experiment and simulation.*

## 5. Conclusions

In this work we studied the undesired influence of diffraction effects and parasitic sub-surface detection in s-SNOM imaging. We show that artefacts may occur in special circumstances due to: (i) edge diffraction, (ii) diffraction on samples harboring structures that act as diffraction gratings (1D and 2D periodic structures) and (iii) generated by buried features. Each case was addressed individually and demonstrated through both experimental methods and simulations. Our study shows that such artefacts tend to affect not only the DC and $1^{st}$ harmonic s-SNOM images, but also higher-harmonics images. For the case of artefacts caused by diffraction due to sample design characteristics, the effect is similar to the undesired case in which the illumination laser beam has a fluctuant intensity. For the case of artefacts associated to signals arising from contrast-active, sub-surface features, s-SNOM has no possibility to discern between surface located features and buried features without some pre-existing knowledge about the investigated sample. Thus, methods to classify the Z-axis origin of s-SNOM signals, e.g. based on Artificial Intelligence, may turn out to be highly useful in the years to come, for enabling s-SNOM imaging with higher fidelity. Overall, this study deepens the comprehension of s-SNOM data and provides a basis for the creation of novel approaches to data acquisition and post-processing. These efforts drive the progress of next-generation s-SNOM imaging and spectroscopy, delivering substantial improvements in resolution and signal-to-noise ratio (SNR).

**Acknowledgements**

This work was supported by the UEFISCDI grant PCE 119 PN-III-P4-PCE-2021-0444 (RESONANO), UEFISCDI grant RO-NO-2019-0601 (MEDYCONAI, NO Grants 2014–2021, Project contract no. 25/2021), and by the UEFISCDI grant PN-III-P2-2.1-PED-2019-2386 (INTEGRAOPTIC). The use of the Neaspec NeaSNOM Microscope was possible due to European Regional Development Fund through Competitiveness Operational Program 2014–2020, Priority axis 1, Project No. P_36_611, MySMIS code 107066 – INOVABIOMED.